\documentclass[english,superscriptaddress,onecolumn]{revtex4-2}
\usepackage[T1]{fontenc}
\usepackage[latin9]{inputenc}
\usepackage{amsmath}
\usepackage{amssymb}
\usepackage{cancel}
\usepackage{graphicx}
\usepackage{babel}
\usepackage{xcolor}
\usepackage{slashed,ulem}

\begin{document}
\title{Vector-Meson Spin Alignment from Anisotropic Quark or Hadron Coalescence}

\author{Wen-Bo Dong}
\email{wenba@mail.ustc.edu.cn}
\affiliation{Department of Modern Physics and Anhui Center for Fundamental Sciences
in Theoretical Physics, University of Science and Technology of China, 
Hefei, Anhui 230026, China}
\affiliation{Institut f\"{u}r Theoretische Physik, Johann Wolfgang Goethe-Universit\"{a}t,
Max-von-Laue-Str.~1, D-60438 Frankfurt am Main, Germany}

\author{Xin-Li Sheng}
\email{sheng@fi.infn.it}
\affiliation{Shanghai Research Center for Theoretical Nuclear Physics, 
NSFC and Fudan University, Shanghai 200438, China}
\affiliation{Universit\`a degli studi di Firenze and INFN Sezione di Firenze,\\
Via G. Sansone 1, I-50019 Sesto Fiorentino (Florence), Italy}
\affiliation{Institut f\"{u}r Theoretische Physik, Johann Wolfgang Goethe-Universit\"{a}t,
Max-von-Laue-Str.~1, D-60438 Frankfurt am Main, Germany}
\affiliation{ExtreMe Matter Institute EMMI,
GSI Helmholtzzentrum f\"{u}r Schwerionenforschung GmbH,
Planckstrasse 1, 64291 Darmstadt, Germany}

\author{Yi-Liang Yin}
\email{yinyiliang@mail.ustc.edu.cn}
\affiliation{Department of Modern Physics and Anhui Center for Fundamental Sciences
in Theoretical Physics, University of Science and Technology of China, 
Hefei, Anhui 230026, China}
\affiliation{Institut f\"{u}r Theoretische Physik, Johann Wolfgang Goethe-Universit\"{a}t,
Max-von-Laue-Str.~1, D-60438 Frankfurt am Main, Germany}

\author{Dirk H.~Rischke}
\email{drischke@itp.uni-frankfurt.de}
\affiliation{Institut f\"{u}r Theoretische Physik, Johann Wolfgang Goethe-Universit\"{a}t,
Max-von-Laue-Str.~1, D-60438 Frankfurt am Main, Germany}
\affiliation{Helmholtz Research Academy Hesse for FAIR, Campus Riedberg,
Max-von-Laue-Strasse 12, D-60438 Frankfurt am Main, Germany}

\author{Qun Wang}
\email{qunwang@ustc.edu.cn}
\affiliation{Department of Modern Physics and Anhui Center for Fundamental Sciences
in Theoretical Physics, University of Science and Technology of China, 
Hefei, Anhui 230026, China}
\affiliation{School of Mechanics and Physics, Anhui University of Science and Technology,\\
Huainan, Anhui 232001, China}

\begin{abstract}
The distribution of particles is highly anisotropic in the initial stage of a heavy-ion collision.
In this paper we demonstrate that this anisotropy induces a sizable effect on the spin alignment of vector mesons. 
We study two different production mechanisms for $\phi$ and $K^{*0}$ mesons, on one hand the coalescence of quarks and on the other that of pseudoscalar mesons. 
In the quark-coalescence picture where $\phi$ and $K^{*0}$ are produced via a bare vector coupling to quarks, a negative $\delta\rho_{00}^y$ of order $10^{-3}$ is observed. 
In contrast, when $\phi$ and $K^{*0}$ are produced via quark coalescence with a vertex with spin-orbit coupling, or when they are produced via pseudoscalar-meson coalescence, a positive $\delta\rho_{00}^y$ emerges. 
In all cases, the magnitude of the spin alignment is directly proportional to the degree of anisotropy.
The sign difference between the cases provides a possibility to clarify the production mechanism for vector mesons. 
\end{abstract}
\maketitle

\section{Introduction}

In non-central heavy-ion collisions (HICs), the large orbital angular momentum carried by the colliding nuclei can be partially transferred to the medium and, via spin-orbit coupling, can give rise to a non-vanishing polarization of particles. 
This phenomenon, known as global spin polarization, was first proposed by Liang and Wang in 2005 \citep{Liang:2004ph} and was experimentally observed in Au+Au collisions by the STAR collaboration in 2017 \citep{STAR:2017ckg}. 
The experimental data show an increasing trend of the global polarization of $\Lambda$ hyperons with collision energy decreasing from 200 to 7.7 GeV. 
The qualitative and quantitative behavior of the global polarization can be well described in a fluid-dynamical framework  \citep{Becattini:2013fla,Becattini:2016gvu,Becattini:2021iol,Fu:2021pok,Pang:2012he,Yi:2021ryh}. 

In addition to spin-1/2 hyperons, spin-1 vector mesons can also be polarized \citep{Liang:2004xn}, characterized by 
a deviation of the spin density matrix $\rho_{\lambda_{1}\lambda_{2}}$ from the unit matrix divided by 3.
In particular, a nonzero value of
$\delta \rho_{00} \equiv \rho_{00} - 1/3$ indicates that the vector meson in the spin-0 state is produced with a probability different from that of the spin-$(\pm1)$ states.
This effect is known as the spin alignment of vector mesons. 
In 2022, the STAR collaboration measured $\delta \rho_{00}$ in the direction transverse to the reaction plane~\footnote{As the reaction plane is commonly taken to be the $(x,z)$-plane, the direction transverse to the reaction plane points in the $y$-direction.}, for both $\phi$ and $K^{*0}$ mesons in Au+Au collisions at RHIC energies \citep{STAR:2022fan}. 
They observed that $\delta \rho_{00}^y$ for the $\phi$ meson is positive for all collision energies in the RHIC beam-energy range.
In contrast, $\delta \rho_{00}^y$ for $K^{*0}$ is within uncertainties consistent with zero in the same range of collision energies. 
The ALICE collaboration observed a negative $\delta \rho_{00}^y$ for low-momentum $\phi$ and $K^{*0}$ mesons produced in Pb+Pb collisions at 2.76 TeV \citep{ALICE:2019aid}. 
These results cannot be explained within existing theoretical approaches, indicating that another mechanism is required.

Recently strong vector-field fluctuations \citep{Sheng:2020ghv,Sheng:2022wsy,Sheng:2022ffb,Kumar:2022ylt,Kumar:2023ghs,Yang:2024qpy} or a thermal-shear contribution \citep{Li:2022vmb,Dong:2023cng,Zhang:2024mhs,Yang:2024fkn,Wang:2025mfz} have been proposed to explain the spin alignment. 
In the model with strong vector-field fluctuations, the strange quark and anti-quark are polarized by the same, and thus highly correlated, vector fields, leading to a non-zero spin alignment of the $\phi$ meson \citep{Sheng:2020ghv,Sheng:2022wsy}.
In contrast, the spin alignment of $K^{*0}$ vanishes because its constituent quark and anti-quark are polarized by different vector fields, which are correlated to a lesser degree \citep{Sheng:2020ghv}. 
Such a mechanism can be tested in experiments through the hyperon spin correlation as proposed in Ref.~\citep{Sheng:2025puj}. 

In this paper, we suggest a new source of spin alignment for vector mesons such as $\phi$ and $K^{*0}$, namely when they are produced via coalescence from quarks or pseudoscalar mesons with anisotropic momentum distributions. 
In the early stages of a heavy-ion collision, the expansion in longitudinal (i.e., beam) direction is much stronger than in the transverse directions, and it takes a finite time for the system to isotropize. 
If the isotropization time is sufficiently long, a large anisotropy can exist even in the final-state distribution. 
As we will show in this paper, this anisotropy generates a non-vanishing spin alignment for vector mesons. 
As representative examples, we choose two production mechanisms: $q\overline{q}$ coalescence and pseudoscalar-meson coalescence, corresponding to production at the quark and hadron level, respectively. 
For the quark coalescence, we consider two types of vertices: a bare vector-interaction vertex \citep{Xu:2019ilh,Xu:2021mju} and a tensor-interaction vertex giving rise to spin-orbit coupling (LS coupling) \citep{Zhao:1998fn}.  
Our results show that the former mechanism will introduce a negative spin alignment of order $10^{-3}$, while the latter gives a positive spin alignment of order $10^{-2}$ independent of the particle species. 

This paper is organized as follows: In Sec.~\ref{sec:anisotropic-hydrodynamics}, we briefly introduce the framework of anisotropic hydrodynamics and the Romatschke-Strickland (RS) distribution~\citep{Florkowski:2010cf,Martinez:2010sc,Alqahtani:2017mhy}. 
Then we apply this distribution to the production of $\phi$ and $K^{*0}$ in HICs and calculate the spin alignment in the above mentioned production processes. 
The results are presented in Sec.~\ref{sec: phi_meson}. 
A discussion, the conclusions, and an outlook are given in Sec.~\ref{sec: conclusion}. 

We adopt the following notations: $g^{\mu\nu}=\text{diag}\left(1,-1,-1,-1\right)$ and  $x^{\mu}=\left(x^{0},\mathbf{x}\right)$ where $\mu,\nu=0,1,2,3$. 
Greek letters are used for tensor indices in four-dimensional space-time, while lowercase Latin letters denote the corresponding spatial components. 
We use natural units $\hbar = c = k_B = 1$.

\section{Anisotropic Fluid Dynamics for Heavy-Ion Collisions\label{sec:anisotropic-hydrodynamics}}
In relativistic heavy-ion collisions, the longitudinal expansion is stronger than the transverse one, which naturally makes the medium anisotropic. 
Such a system can be described by anisotropic hydrodynamics (aHydro) \citep{Florkowski:2010cf,Martinez:2010sc,Alqahtani:2017mhy}.
Here, the particle current and the energy-momentum tensor are constructed in accordance with the residual symmetry of the system~\citep{Molnar:2016vvu,Rocha:2023ilf}. 
A commonly used simplification is to assume that the system consists of an ideal fluid with an anisotropic pressure, for which the energy-momentum tensor takes the form
\begin{equation}\label{eq:Tmunu}
T^{\mu\nu} =(\varepsilon+P_T) u^{\mu}u^{\nu}-g^{\mu\nu}P_{T}-\left(P_{T}-P_{L}\right)l^{\mu}l^{\nu}\;.
\end{equation}
Here, $u^\mu$ is the fluid-velocity vector and $l^{\mu}$ is a space-like unit vector orthogonal to $u^\mu$ specifying the direction of anisotropy. 
For a purely longitudinal boost-invariant expansion and assuming the anisotropy to point in beam direction, we have $u^\mu = (t, 0,0,z)/\tau$ and $l^\mu= (z,0,0,t)/\tau$, where $\tau = \sqrt {t^2-z^2}$.
In this case, $P_L$ and $P_T$ denote the pressure in the longitudinal and transverse direction, respectively.
The anisotropy is then quantified by the ratio $P_L/P_T$, which is usually smaller than 1 according to aHydro calculations \citep{Molnar:2016vvu,Alqahtani:2017jwl}. 
Finally, $\varepsilon$ in Eq.~\eqref{eq:Tmunu} is the energy density.

\subsection{Romatschke-Strickland distribution}

Following Romatschke and Strickland (RS), the anisotropic momentum distribution for massless fermions (bosons) can be parameterized as \citep{Romatschke:2003ms}
\begin{eqnarray}\label{eq:RS-distribution}
f^{\text{RS}}(p,\xi) & = & \frac{d_s}{\exp\left[\beta\sqrt{p^{\mu}p^{\nu}(u_\mu u_\nu +\xi l_\mu l_\nu)}\right]\pm1}\;,
\end{eqnarray}
where $\beta=1/T$ is the inverse temperature and the parameter $\xi$ characterizes the magnitude of the momentum anisotropy. 
Here, $d_s$ denotes the degeneracy of a momentum state, corresponding to the number of internal degrees of freedom, such as spin, flavor, and color. 
For $\xi>0$ ($0>\xi>-1$), $f^\text{RS}$ corresponds to a Fermi-Dirac/Bose-Einstein distribution with oblate (prolate) spheroidal deformation along the $z$-axis. 
Using the RS ansatz, the energy-momentum tensor $T^{\mu\nu}$ is calculated via 
\begin{equation}
T^{\mu\nu}_\text{RS}(\xi)=\int \frac{d^{3}\mathbf{p}}{(2\pi)^3\, |\mathbf{p}|}\, p^{\mu}p^{\nu}\,f^{\text{RS}}(p,\xi)\;.
\end{equation}
Substituting Eq.~\eqref{eq:RS-distribution} into the above equation and parameterizing the result as shown in Eq.~\eqref{eq:Tmunu}, we obtain the energy density and the pressure components as
\begin{align}
\varepsilon_{\text{RS}}(\xi) & =\frac{1}{2}\left(\frac{1}{1+\xi}+\frac{\arctan\sqrt{\xi}}{\sqrt{\xi}}\right)\varepsilon_{\text{iso}}\equiv\mathcal{R}\left(\xi\right)\varepsilon_{\text{iso}}\;,\nonumber \\
P_{T}^{\text{RS}}(\xi) & =\frac{3}{2\xi}\,\frac{1+\left(\xi^{2}-1\right)\mathcal{R}\left(\xi\right)}{\xi+1}\, P_{\text{iso}}\;,\nonumber \\
P_{L}^{\text{RS}}(\xi) & =\frac{3}{\xi}\,\frac{\left(\xi+1\right)\mathcal{R}\left(\xi\right)-1}{\xi+1}\,P_{\text{iso}}\;.\label{eq:PL,PT}
\end{align}
where $\varepsilon_\text{iso}\equiv\varepsilon_{\text{RS}}(0)$ and $P_\text{iso}\equiv P_{T}^{\text{RS}}(0)=P_{L}^{\text{RS}}(0)$ denote the corresponding values in the isotropic limit. 

\begin{figure}
    \centering
    \includegraphics[width=0.5\linewidth]{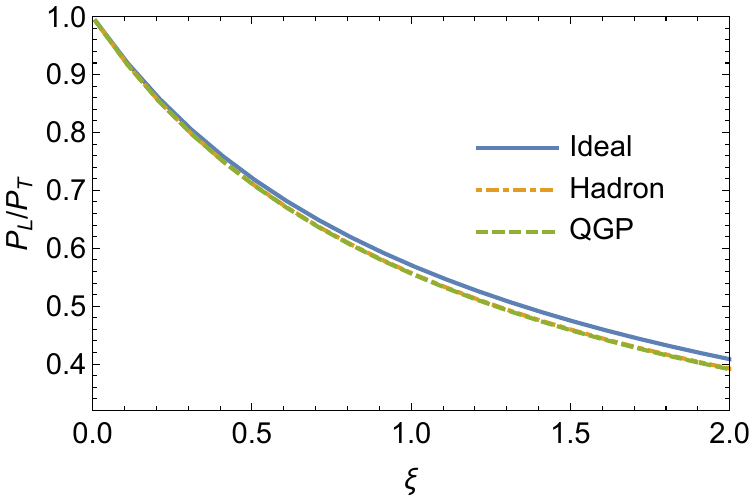}
    \caption{The pressure ratio $P_{L}/P_{T}$ as a function of the anisotropy parameter $\xi$ in a hadron gas (dot-dashed line), in a QGP (dashed line), and in the ideal case with massless particles (solid line). The dot-dashed and dashed lines almost coincide.}
    \label{fig:PLPT_xi}
\end{figure}

In this work, we study the formation of vector mesons through the coalescence of quarks or pseudoscalar mesons at the freeze-out stage. 
The considered particles are in general massive ones. 
We therefore generalize the RS distribution in Eq.~\eqref{eq:RS-distribution} to the case of particles with non-zero mass $m_i$, \citep{Romatschke:2003ms,Bazow:2013ifa}
\begin{align}\label{eq:RS-dis-massive}
f^{\text{RS}}_i(p,\xi) & =\frac{d_s}{\exp\left[\beta\sqrt{p_{T}^{2}+(\xi+1)p_{z}^{2}+m^{2}_i}\right]\pm1}\;.
\end{align}
The total energy-momentum tensor is the sum over all particle species,
\begin{equation}
T^{\mu\nu}(\xi)=\sum_i \int \frac{d^{3}\mathbf{p}}{(2\pi)^3\sqrt{{\bf p}^2+m_i^2}}\, p^{\mu}p^{\nu}\,f_i^{\text{RS}}(p,\xi)\;.
\end{equation}
In Fig.~\ref{fig:PLPT_xi} we show the pressure ratio $P_L/P_T$ as a function of the anisotropy parameter $\xi$ for massless particles (blue solid line), a hadron gas (yellow dot-dashed line), and a quark-gluon plasma (QGP) (green dashed line) at $T=150$ MeV.
For the hadron gas model, pions, kaons, protons, and neutrons are taken into account, while heavier hadrons are neglected. 
For the QGP, we assume that the masses of $u$ and $d$ quarks, as well as of gluons are modified by the presence of a thermal medium of temperature $T$, i.e., $m_{u,d}=gT/\sqrt{6},\,m_g=(\sqrt{3}/2)gT$. 
We take a value of $g=2.3$ for the strong coupling constant.
The $s$ quark mass is chosen as $m_s=420$ MeV. 
The differences between the three cases are relatively small, indicating that the relationship between $\xi$ and $P_L/P_T$ is nearly model-independent. 
According to the analytical results for massless particles in Eq.~(\ref{eq:PL,PT}), the ratio reads 
\begin{equation}
\label{eq:PL/PT_xi}
\frac{P_L}{P_T}=2\, \frac{(\xi+1)\mathcal{R}(\xi)-1}{1+(\xi^2-1)\mathcal{R}(\xi)}\;,
\end{equation}
which reduces to 1 in the isotropic limit $\xi\rightarrow0$.

\subsection{Anisotropy in Bjorken expansion}

Now we try to estimate the anisotropy in the simple case of the time-honored Bjorken expansion scenario~\citep{Bjorken:1982qr}. 
Taking into account the boost invariance in the longitudinal direction and neglecting the expansion in the transverse directions, the conservation equation $\partial_\mu T^{\mu\nu}_\text{RS}=0$  simplifies to \citep{Martinez:2010sc}
\begin{eqnarray}
\frac{1}{1+\xi}\partial_{\tau}\xi-\frac{6}{\gamma}\partial_\tau \gamma& = & \frac{2}{\tau}+2\Gamma\left[1-\mathcal{R}^{3/4}\left(\xi\right)\sqrt{1+\xi}\right]\;,\nonumber \\
\frac{\mathcal{R}^{\prime}\left(\xi\right)}{\mathcal{R}\left(\xi\right)}\partial_{\tau}\xi+\frac{4}{\gamma}\partial_\tau \gamma & = & -\frac{1}{\tau}\left[1+\frac{1}{\xi}-\frac{1}{\xi\left(\xi+1\right)\mathcal{R}\left(\xi\right)}\right]\;,
\end{eqnarray}
where $\gamma$ is related to the temperature as $T=\gamma\mathcal{R}^{1/4}(\xi)$, $\Gamma$ denotes the relaxation rate of the system, which is proportional to the ratio of the entropy density $s$ to the shear viscosity $\eta$, i.e, $\Gamma\propto s/\eta$. 


A detailed numerical investigation of these equations has been performed in Ref.~\citep{Martinez:2010sd}. 
The results show that $P_{L}/P_{T}$ on the freeze-out hypersurface increases as $\eta/s$ decreases and approaches 1 in the ideal-fluid limit. 
To illustrate this relation explicitly, we take Navier-Stokes (NS) theory as an approximation. 
In NS theory, the difference between $P_{L}$ and $P_{T}$ is induced by the shear-stress tensor $\pi^{\mu\nu}$, whose non-zero components in the local rest frame are $\pi_{xx}=\pi_{yy}=-\pi_{zz}/2=2\eta/(3\tau)$. 
Combining this with the ultrarelativistic ideal gas equation of state $\varepsilon=3P=(3/4)Ts$, one obtains
\begin{eqnarray}
\left(\frac{P_{L}}{P_{T}}\right)_{\text{NS}} & = & \frac{3T\tau-16\eta/s}{3T\tau+8\eta/s}\;.\label{eq:PL/PT_NS}
\end{eqnarray}
A simple quantitative estimate of the ratio~(\ref{eq:PL/PT_NS}) can be obtained as follows.
At high collision energies, $\eta/s$ can be approximated by the KSS bound $1/(4\pi)$~\citep{Kovtun:2004de}. 
For a typical freeze-out time $\tau=10$ fm/c \citep{Schafer:2021csj} and temperature $T=150$ MeV, we obtain from Eq.~\eqref{eq:PL/PT_NS} the ratio $P_{L}/P_{T}\approx0.92$.
From Eq.~\eqref{eq:PL/PT_xi} we then deduce $\xi\approx0.1$. 
At low collision energies, $\eta/s\sim0.2$  \citep{Auvinen:2017pny} and the typical freeze-out time is usually shorter, e.g., $\tau\approx8$ fm/c  \citep{Schafer:2021csj}. The ratio \eqref{eq:PL/PT_NS} then decreases to $P_{L}/P_{T}\approx0.77$, implying from Eq.~\eqref{eq:PL/PT_xi} a sizable anisotropy $\xi\approx0.4$. 
A more detailed investigation~\citep{Molnar:2016gwq} also shows a sizable $P_{L}/P_{T}$ in the large-$\eta/s$ regime. 
A more precise estimate of $P_L/P_T$ as a function of collision energy can be obtained from a numerical simulation, which will be discussed in a forthcoming paper. 

\section{Spin alignment of $\phi$ and $K^{*0}$ mesons from anisotropic distributions \label{sec: phi_meson}}

In this section, we  calculate the spin alignment of $\phi$ and $K^{*0}$ mesons produced via the coalescence process of quarks and pseudoscalar mesons, assuming that the coalescing particles have anisotropic momentum distributions. 

\subsection{General expression}

Before presenting an explicit calculation, we give a general formula for the spin alignment of the vector meson. 
Considering the production of a vector meson through a coalescence process $2\rightarrow 1$, the cross section can be put into the general form 
\begin{eqnarray}\label{eq:G_sigma}
\sigma_{\lambda}\left(p,q\right) & = & \epsilon_{\mu}^{*}\left(p,\lambda\right)\epsilon_{\nu}\left(p,\lambda\right)\mathcal{M}^{\mu\nu}\left(p,q\right)\;,
\end{eqnarray}
where $p$ and $q$ denote the total and relative momentum of two incoming particles, respectively. 
Due to momentum conservation, the momentum of the outgoing vector meson is equal to $p$. 
Moreover, $\epsilon_\mu(p,\lambda)$ denotes the polarization vector of the vector meson with spin projection $\lambda$ in the quantization direction in the lab frame. 
This vector can be obtained via a Lorentz boost from the polarization vector $(0,\boldsymbol{\epsilon}_{\lambda})$ in the rest frame of the particle
\begin{align}
\epsilon^{\mu}\left(\lambda,p\right) & =\left(\frac{\mathbf{p}\cdot\boldsymbol{\epsilon}_{\lambda}}{m_V},\boldsymbol{\epsilon}_{\lambda}+\frac{\mathbf{p}\cdot\boldsymbol{\epsilon}_{\lambda}}{m_V\left(E_{p}+m_V\right)}\mathbf{p}\right)\;,
\end{align}
where $m_V$ is the mass of the vector meson and $E_p \equiv \sqrt{\mathbf{p}^2 + m_V^2}$ is its on-shell energy. 
If the spin quantization direction is chosen as the $y$-axis, the vector $\boldsymbol{\epsilon}_{\lambda}$ takes the following form
\begin{align}
\boldsymbol{\epsilon}_{0} & =\left(0,1,0\right)\;,\nonumber \\
\boldsymbol{\epsilon}_{\pm1} & = \mp \frac{1}{\sqrt{2}}\left(\pm i,0,1\right)\;.
\label{eq:epsilon_rest}
\end{align}
Since $\mathcal{M}^{\mu\nu}$ in Eq.~(\ref{eq:G_sigma}) is a hermitian Lorentz tensor that depends on $p^\mu$ and $q^\mu$, its general expansion reads,
\begin{equation}\label{eq:expand-M-munu}
\mathcal{M}^{\mu\nu}(p,q)=c_1\,m_{V}^2\,g^{\mu\nu}+c_2\,q^\mu q^\nu+c_3\,p^\mu p^\nu+c_4\,(p^\mu q^\nu+q^\mu p^\nu)
+ ic_5 (p^\mu q^\nu-q^\mu p^\nu) + ic_6 \,\epsilon^{\mu \nu \alpha \beta} p_\alpha q_\beta\;,
\end{equation}
where $c_i$ are real dimensionless coefficients that depend on the Lorentz invariants $p^2$, $q^2$, and $p\cdot q$. 
Here we have neglected the contribution from the hot and dense medium to $\mathcal{M}^{\mu\nu}$, which singles out a particular rest frame and renders the decomposition much more complicated. 
Applying the constraint $p\cdot\epsilon\left(p\right)=0$, Eq.~(\ref{eq:G_sigma}) can be simplified to
\begin{eqnarray}
\sigma_{\lambda}(p,q) & = & -c_1\,m_{V}^2+c_{2}\left|q\cdot\epsilon\left(p,\lambda\right)\right|^{2}
+ i c_6 \, \epsilon^{\mu \nu \alpha \beta} \epsilon_\mu^*(p,\lambda) \epsilon_\nu(p,\lambda) p_\alpha q_\beta\;.\label{eq:G cross section}
\end{eqnarray}
The production rate for the spin-$\lambda$ state is then given by 
\begin{align}
\Gamma_\lambda &\propto\int\frac{d^3{\bf p}\,d^3{\bf q}}{4E_{p_1}E_{p_2}}\delta(E_p-E_{p_1}-E_{p_2})\;
\sigma_{\lambda}(p,q)\;f_1(p_1)f_2(p_2)\;,
\label{eq:Gamma_lambda}
\end{align}
where $p_{1,2}=(p \pm q)/2$ are the momenta of two incoming particles, $f_i(p_i)$ ($i=1,2$) are their distribution functions, and $E_p$ and $E_{p_i}$ ($i=1,2$) are the energies of the outgoing meson and incoming particles, respectively. 
Taking a sum over all polarization degrees of freedom, $\lambda=0,\pm 1$, we obtain the total production rate
\begin{equation}\label{eq:Gamma_lambda_tot}
\sum_{\lambda=0,\pm1}\Gamma_\lambda\propto\int\frac{d^3{\bf p}\,d^3{\bf q}}{4E_{p_1}E_{p_2}}\delta(E_p-E_{p_1}-E_{p_2})\left\{-3\,c_1\,m_{V}^2+c_{2}\left[\frac{(p\cdot q)^2}{m_{V}^2}-q^2\right]\right\}f_1(p_1)f_2(p_2)\;.
\end{equation}
The spin alignment of the vector meson is then evaluated as
\begin{equation}\label{eq:delta-rho-00}
\delta\rho_{00}^y(p)\equiv\rho_{00}^y(p)-\frac13=\frac{d\Gamma_0/d^3{\bf p}}{\sum_{\lambda=0,\pm1}d\Gamma_\lambda/d^3{\bf p}}-\frac13\;,
\end{equation}
where the differential production rate $d\Gamma_0/d^3{\bf p}$ can be obtained by removing the integral over $\mathbf{p}$ in Eq.~(\ref{eq:Gamma_lambda}), while $d\Gamma_\lambda/d^3{\bf p}$ summed over all polarizations can be obtained by removing the integral over $\mathbf{p}$ in Eq.~(\ref{eq:Gamma_lambda_tot}). 
If the produced meson is non-relativistic, i.e., $|{\bf p}|\ll m_V$, the spin alignment can be estimated by
\begin{equation}\label{eq:G rho00-1/3}
\delta\rho_{00}^y(p)\approx \frac{\left\langle q_y^2\right\rangle - \left\langle{\bf q}^2\right\rangle /3}{\left\langle{\bf q}^2\right\rangle-3\,  m_{V}^2\, c_1/c_2}\;,
\end{equation}
where  
\begin{equation} \label{eq:average}
\langle A \rangle\equiv \frac{\int d^3{\bf q}/(E_{p_1}E_{p_2})\delta(E_p-E_{p_1}-E_{p_2})\,A\,f_1(p_1)f_2(p_2)}{\int d^3{\bf q}/(E_{p_1}E_{p_2})\delta(E_p-E_{p_1}-E_{p_2})f_1(p_1)f_2(p_2)}
\end{equation}
denotes the average of the quantity $A$ over the relative momentum $\mathbf{q}$ 
and depends on the vector-meson momentum $\mathbf{p}$. 
For isotropic distribution functions, the numerator in Eq.~(\ref{eq:G rho00-1/3}) vanishes because of symmetry.
In contrast, for anisotropic distribution functions, $\left\langle q_y^2\right\rangle\neq\left\langle {\bf q}^2\right\rangle/3$, which leads to a non-vanishing $\delta\rho_{00}$.
This holds not only in the non-relativistic limit, but also for the general case as long as $c_2\neq 0$. 
However, the sign and magnitude of $\delta\rho_{00}$ depends on the values of
$c_1$ and $c_2$, as specified by the underlying interaction vertex. 

In the following subsections, we consider two scenarios of vector-meson production: pseudoscalar-meson coalescence and $q\bar{q}$ coalescence, corresponding to production at the hadron and quark level, respectively. 
In the latter case we also study two different interaction vertices, one corresponding to a pure vector interaction, and the other to a spin-orbit interaction.
The anisotropy is included through the generalized RS distribution (\ref{eq:RS-dis-massive}) controlled by the anisotropy parameter $\xi$.

\subsection{Pseudoscalar-meson coalescence\label{subsec:coalesence}}

\begin{figure}
\begin{centering}
\includegraphics[scale=0.75]{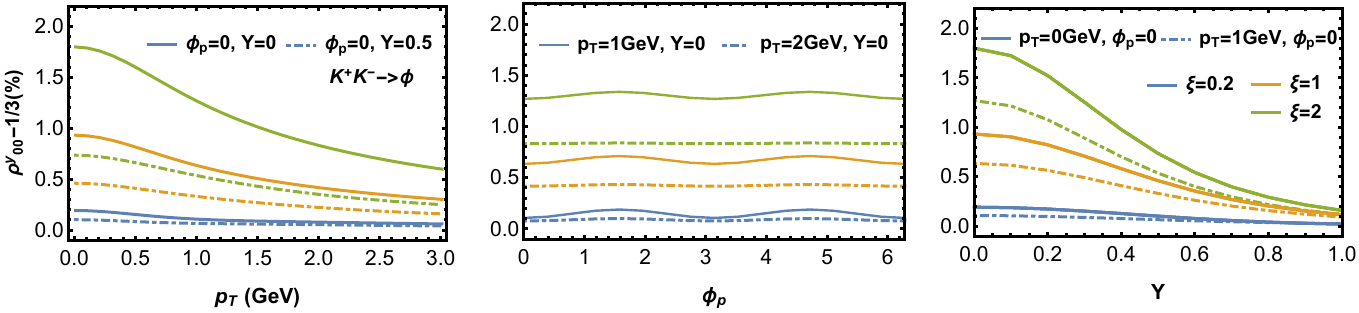}
\includegraphics[scale=0.75]{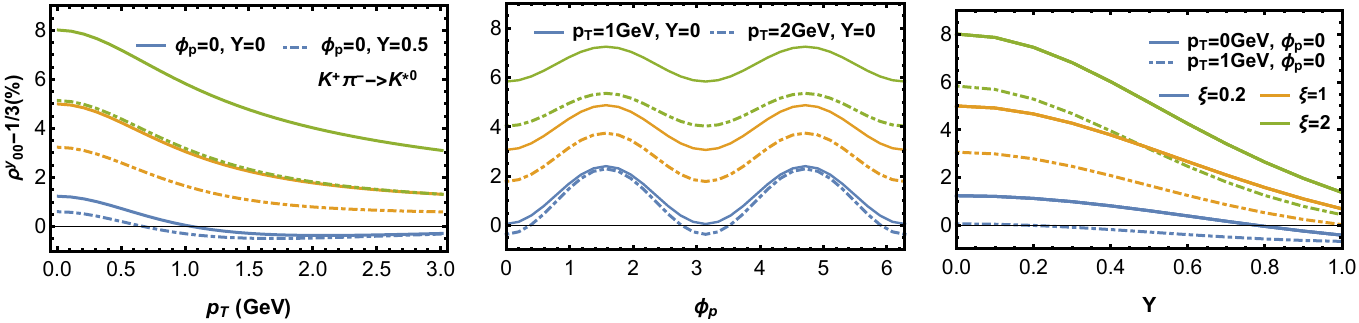}
\par\end{centering}
\caption{The spin alignment $\rho_{00}^y-1/3$ as function of $p_{T}$ (left), $\phi_{p}$ (middle), and $Y$ (right) for $\phi$ (upper row) and $K^{*0}$ (lower row) mesons. 
The blue, orange, and green lines correspond to anisotropy parameters $\xi=0.2,\,1$, and $2$, respectively. \label{fig:KK_xi}}
\end{figure}

In this subsection, we study the spin alignment induced by anisotropy in a pseudoscalar-meson coalescence process. 
The dominant hadronic channels for the production of $\phi$ and $K^{*0}$, i.e., $K+K\rightarrow \phi$ and $K+\pi\rightarrow K^{*0}$, have been studied for decades and are well described by chiral perturbation theory \citep{Ecker:1988te,Scherer:2002tk}. 
The corresponding interaction Lagrangian at leading order reads
\begin{align}
\mathcal{L}_{\phi KK} & =g_{\phi KK}\phi^{\mu}\left(K^{+}\partial_{\mu}K^{-}-K^{-}\partial_{\mu}K^{+}+K^{0}\partial_{\mu}\overline{K}^{0}-\overline{K}^{0}\partial_{\mu}K^{0}\right)\;,\nonumber\\
\mathcal{L}_{K^{*0}K\pi}&=g_{K^{*0}K\pi}K^{*0}_{\mu}\left(K^{-}\partial^\mu\pi^{+}-\pi^{+}\partial^\mu K^{-}+K^0\partial^\mu\pi^0-\pi^0\partial^\mu K^0\right)\;.\label{eq:phi_KK_int}
\end{align}
Here, the coupling constants $g_{\phi KK}$ and $g_{K^{*0}K\pi}$ can be extracted from the decay widths of the respective vector mesons. 
They will eventually cancel out when taking the ratio in Eq.~(\ref{eq:delta-rho-00}), thus we do not present them explicitly in this paper. 
Due to isospin symmetry, the coalescence process of charged particles is identical to that of neutral particles. 
We therefore focus on the channels $K^{+}+K^{-}\rightarrow\phi$ and $K^{+}+\pi^{-}\rightarrow K^{*0}$. From the Lagrangian given in Eq.~(\ref{eq:phi_KK_int}),  the coefficients $c_1$ and $c_2$ associated with Eq.~(\ref{eq:phi_KK_int}) are given by
\begin{align}
c_1^{\phi}=0\;,\quad\quad &c_2^{\phi}=2g_{\phi KK}^2\;,\nonumber\\
c_1^{K^{*0}}=0\;,\quad\quad &c_2^{K^{*0}}=2g_{K^{*0} K\pi}^2\;.
\label{c1-c2-meson}
\end{align}
The spin alignments of $\phi$ and $K^{*0}$ mesons can then be calculated from Eqs.~(\ref{eq:Gamma_lambda}), (\ref{eq:Gamma_lambda_tot}), and (\ref{eq:delta-rho-00}), where $f_1$ and $f_2$ are the distribution functions of the pseudoscalar mesons, which are assumed to take the anisotropic form  (\ref{eq:RS-dis-massive}) with $d_s=1$ and $m=m_{K,\pi}$. 
Choosing the physical mass of the vector and pseudoscalar mesons, $m_K=494$ MeV, $m_\pi=140$ MeV, $m_\phi=1020$ MeV, and $m_{K^{*0}}=892$ MeV, and taking a typical freeze-out temperature of $T=150$ MeV, we numerically calculate the spin alignments of  $\phi$ and $K^{*0}$ mesons as functions of the transverse momentum $p_T$, the azimuthal angle $\phi_p$, and the rapidity $Y$. 
Here, the three-momentum of the vector meson is expressed as $\mathbf{p}=(p_T\cos \phi_p, p_T\sin \phi_p, \sqrt{p_T^2+m_V^2}\sinh Y)$. 
As shown in Fig.~\ref{fig:KK_xi}, the spin alignment varies with $\phi_p$ in a cosine-like pattern and decreases with increasing $p_T$ or $Y$. 
We find that vector mesons produced from hadron coalescence have a positive $\delta \rho_{00}^y$, and the magnitude increases with $\xi$. 
For $\phi$ and $K^{*0}$ mesons with the same momentum and $\xi$, the spin alignment of $K^{*0}$ is 4-5 times larger than that of $\phi$, which arises from the mass difference between the pion and kaon.

\subsection{$q\overline{q}$ coalescence with bare vector coupling \label{subsec:coalesence-process}}

For the quark-antiquark coalescence process, the interaction Lagrangian corresponding to a bare vector coupling reads \citep{Xu:2019ilh,Xu:2021mju}
\begin{align}
\mathcal{L}_{\phi} & =ig_{\phi s\bar{s}}\overline{\psi}_s\gamma^{\mu}\psi_s\phi_{\mu}\;,\nonumber\\
\mathcal{L}_{K^{*0}} & =ig_{K^{*0}d\bar{s}}\overline{\psi}_d\gamma^{\mu}\psi_{s}K^{*0}_{\mu}\;,
\label{eq:s_sbar_phi}
\end{align}
where $\psi_{d,s}$ are the Dirac spinors of $d$ and $s$ quark, and $g_{\phi s\bar{s}}$ and $g_{K^{*0}d\bar{s}}$ denote the effective coupling constants for the $\phi$-$s$-$\bar{s}$ and $K^{*0}$-$d$-$\bar{s}$ vertices, respectively. 
Following a calculation similar to that in Sec.~\ref{subsec:coalesence}, the coefficients are given by
\begin{align}
c^{\phi}_1=-2g_{\phi s\bar{s}}^2\;, \quad\quad & c^{\phi}_2=-2g_{\phi s\bar{s}}^2\;, \nonumber\\
c_1^{K^{*0}}=-2g_{K^{*0}d\bar{s}}^2\left[1-\frac{(m_s-m_d)^2}{m_{K^{*0}}^2}\right]\;, \quad\quad &c_2^{K^{*0}}=-2g_{K^{*0}d\bar{s}}^2\;.
\label{eq:c12_ssbar}
\end{align}
Choosing a temperature of $T=150$ MeV and setting the $d$-quark mass to $m_d=300$ MeV, we numerically calculate the spin alignment of $\phi$ and $K^{*0}$ for different choices of the $s$-quark mass $m_s$ and the anisotropy parameter $\xi$.

\begin{figure}
\begin{centering}
\includegraphics[scale=0.75]{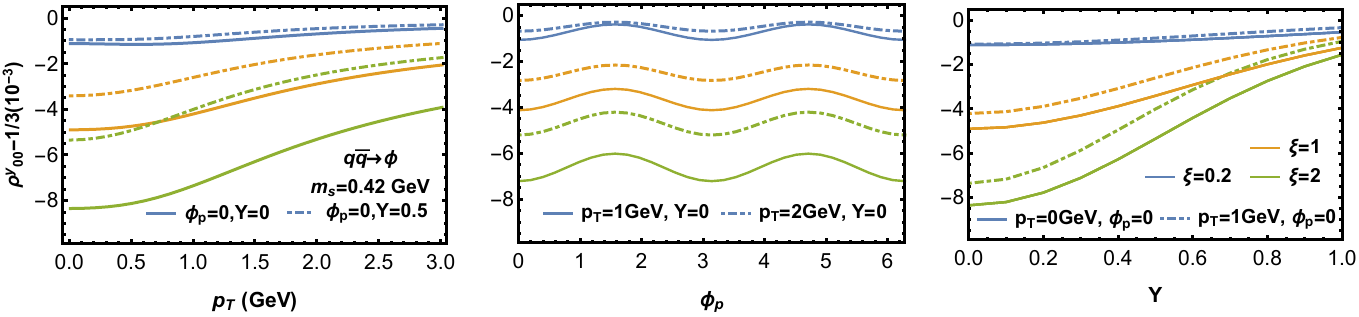}
\includegraphics[scale=0.75]{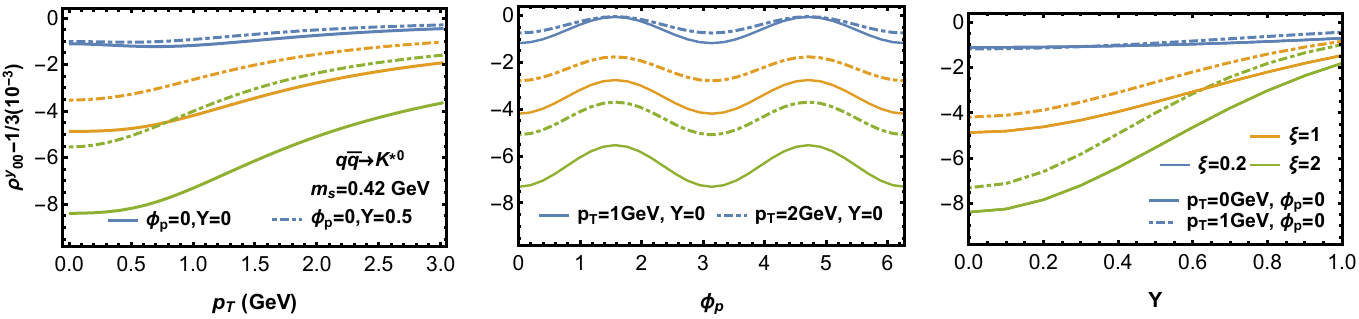}
\par\end{centering}
\caption{Similar to Fig.~\ref{fig:KK_xi} but for $\phi$ and $K^{*0}$ mesons produced by quark coalescence with a bare vector coupling. \label{fig:xi_dependent}}
\end{figure}

\begin{figure}
\begin{centering}
\includegraphics[scale=0.75]{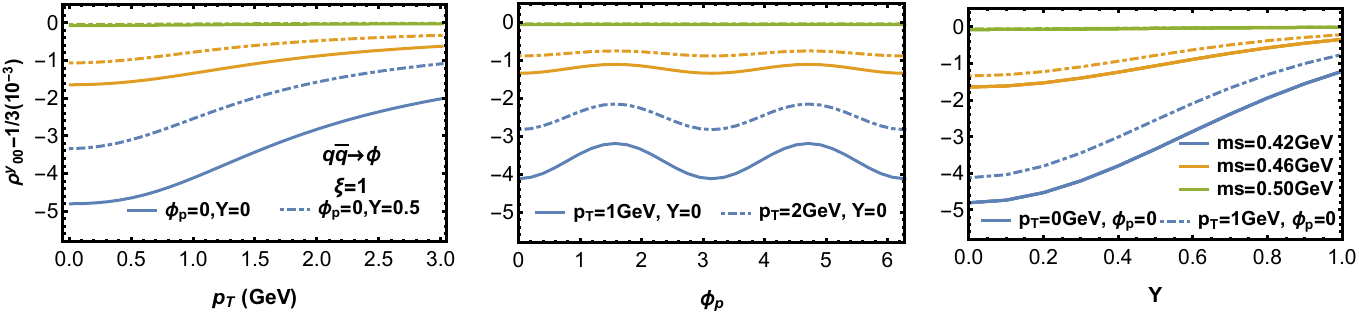}
\includegraphics[scale=0.75]{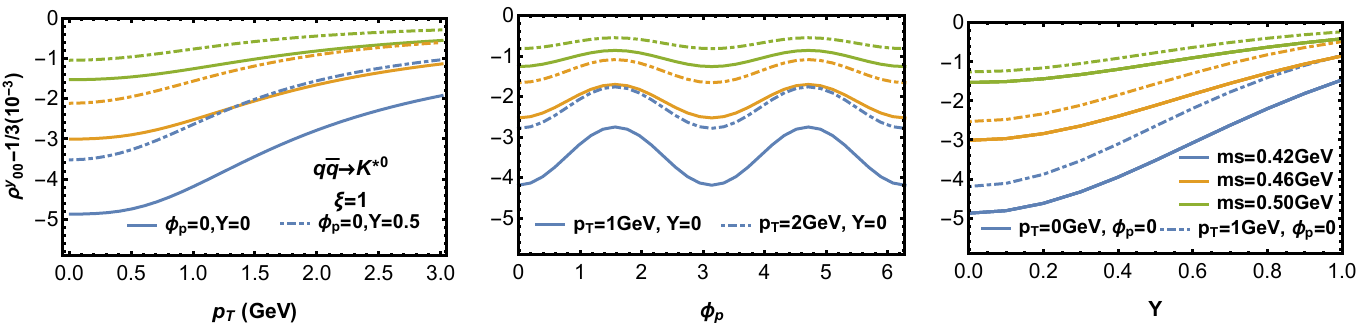}
\par\end{centering}
\caption{Same as Fig.~\ref{fig:xi_dependent} but with $\xi=1$ and $m_s=0.42,0.46,0.5$ GeV. \label{fig:ms-dependent}}
\end{figure}

As shown in Figs.~\ref{fig:xi_dependent} and \ref{fig:ms-dependent}, $\delta \rho_{00}^y$ for $\phi$ and $K^{*0}$ is negative. 
Their dependence on $p_T$, $\phi_p$, and $Y$ is similar because the $\phi$-$s$-$\bar{s}$ interaction vertex is similar to that of the $K^{*0}$-$d$-$\bar{s}$ vertex, cf.~Eq.~(\ref{eq:s_sbar_phi}).  
Furthermore, comparing the results for different values of $\xi$ and $m_s$, we conclude that $\delta \rho_{00}^y$ increases in magnitude with $\xi$ and decreases in magnitude with $m_s$. 
The increase with $\xi$ can be easily explained by Eq.~\eqref{eq:G rho00-1/3}: Increasing $\xi$ increases the numerator, as the deviation of $\langle q_y^2\rangle$ from $\langle \mathbf{q}^2\rangle/3$ becomes larger. 
Since the denominator is negative for the quark coalescence picture, this results in an overall negative $\delta \rho_{00}^y$.
The increase of $|\delta \rho_{00}^y|$ with decreasing $m_s$ is less obvious.
Let us for the sake of simplicity consider the spin alignment of $\phi$ mesons.
One can then convince oneself using Eq.~\eqref{eq:average} that, in Boltzmann approximation and to leading order in $\xi$, the following approximation holds in the rest frame of the $\phi$ meson,
\begin{equation}
    \langle q_y^2\rangle-\frac{\langle \mathbf{q}^2\rangle}{3} \sim\frac{\xi m_\phi^3}{T}\left(1-\frac{4 m_s^2}{m_\phi^2} \right)^2\;.
\end{equation}
Therefore, a smaller quark mass leads to a larger $\langle q_y^2\rangle-\langle \mathbf{q}^2\rangle/3$, resulting in a larger $|\delta\rho^y_{00}|$.
The above formula also explains why the spin alignment is very sensitive to small changes of the $s$-quark mass in the range of half the $\phi$-meson mass.
We remark that for the highest anisotropy considered here ($\xi=2$), the ratio $P_{L}/P_{T}\approx 0.4$ is unrealistically small for the typical freeze-out conditions in heavy-ion collisions. 

\subsection{$q\bar{q}$ coalescence with LS coupling vertex\label{sec:LS}}

When the quark mass is small, the vertex in Eq.~(\ref{eq:s_sbar_phi}) no longer dominates the production of vector mesons. 
In this case, the bare vector-coupling vertex should be generalized \citep{Maris:1999nt}. 
One typical choice is $\gamma^\mu+i\zeta \sigma^{\mu\nu} (p_{1\nu}-p_{2\nu})/(2m_q)$, where $\zeta$ represents the ratio of the tensor to the vector coupling \citep{Zhao:1998fn}. 
The additional tensor coupling arises from the spin-orbit (LS) interaction and is usually suppressed by the ratio of the relative momentum to the quark mass, which becomes non-negligible in the small-mass limit. 

More concretely, we generalize the interaction Lagrangian  (\ref{eq:s_sbar_phi}) to the following form \citep{Zhao:1998fn},
\begin{align}
\mathcal{L}_{\phi} & =ig_{\phi s\bar{s}}\overline{\psi}_s\left[\gamma^{\mu}+\frac{\zeta_\phi}{2M}\sigma^{\mu\nu}(\overrightarrow{\partial_\nu}+\overleftarrow{\partial_{\nu}})\right]\psi_s\phi_{\mu}\;,\nonumber\\
\mathcal{L}_{K^{*0}} & =ig_{K^{*0}d\bar{s}}\overline{\psi}_d\left[\gamma^{\mu}+\frac{\zeta_K}{2M}\sigma^{\mu\nu}(\overrightarrow{\partial_\nu}+\overleftarrow{\partial_{\nu}})\right]\psi_{s}K^{*0}_{\mu}\;,
\label{ls-coupling}
\end{align}
where $\overrightarrow{\partial_\nu}$ and $\overleftarrow{\partial_{\nu}}$ act only on the Dirac fields and $\zeta_\phi$ and $\zeta_K$ are the ratios of the tensor to vector coupling for the $\phi$-$s$-$\bar{s}$ and $K^{*0}$-$d$-$\bar{s}$ vertices, respectively. 
The parameter $M$ is a typical mass scale for the tensor interaction part, which will be set to 300 MeV.
Since $\zeta_\phi$ and $\zeta_K$ are free parameters, the precise choice for $M$ is not relevant.

With the help of the Dirac equation, we can convert the Lagrangian to the equivalent form
\begin{align}\label{eq:int-vector-tensor}
\mathcal{L}_{\phi} & =g_{\phi s\bar{s}}\overline{\psi}_s\left[i\gamma^{\mu}\left(1+\zeta_\phi\frac{m_s}{M}\right)+\frac{\zeta_\phi}{2M}\overset{\leftrightarrow}{\partial}^\mu\right]\psi_s\phi_{\mu}\;,\nonumber\\
\mathcal{L}_{K^{*0}} & =g_{K^{*0}d\bar{s}}\overline{\psi}_d\left[i\gamma^{\mu}\left(1+\zeta_K\frac{m_d+m_s}{2M}\right)+\frac{\zeta_K}{2M}\overset{\leftrightarrow}{\partial}^\mu\right]\psi_{s}K^{*0}_{\mu}\;,
\end{align}
where $\overset{\leftrightarrow}{\partial}_\mu\equiv\overrightarrow{\partial_\mu}-\overleftarrow{\partial_{\mu}}$ again act only on the Dirac fields. 
From the Lagrangian, we can obtain the coefficients $c_1$ and $c_2$ that are related to the spin alignment,
\begin{align}
&c^{\phi}_1=-2g_{\phi s\bar{s}}^2\left(1+\zeta_\phi\frac{m_s}{M}\right)^2\;, \nonumber\\  &c^{\phi}_2=-2g_{\phi s\bar{s}}^2\left(1-\zeta_\phi^2\frac{m_s^2}{M^2}-\zeta_\phi^2\frac{m_\phi^2-4m_s^2}{4M^2}\right)\;, \nonumber\\
&c_1^{K^{*0}}=-2g_{K^{*0} d\bar{s}}^2\left(1+\zeta_K\frac{m_d+m_s}{2M}\right)^2\left[1-\frac{(m_d-m_s)^2}{m_{K^{*0}}^2}\right]\;, \nonumber\\ 
&c_2^{K^{*0}}=-2g_{K^{*0} d\bar{s}}^2\left[1-\zeta_K^2\frac{(m_d+m_s)^2}{4M^2}-\zeta_K^2\frac{m_{K^{*0}}^2-(m_d+m_s)^2}{4M^2}\right]\;.
\label{eq:c12_ssbar_LS}
\end{align}
It is obvious that the Lagrangian (\ref{eq:int-vector-tensor}) reduces to Eq.~(\ref{eq:s_sbar_phi}) when $\zeta_\phi=\zeta_K=0$. 
The LS coupling becomes important when the ratios $\zeta_\phi/(1+\zeta_{\phi}m_s/M)$ and $\zeta_K/[1+\zeta_K(m_s+m_d)/2M]$ are sufficiently large. 
In particular, the pure vector couplings in Eq.~(\ref{eq:int-vector-tensor}) vanish when $\zeta_\phi=-M/m_s$ and $\zeta_K=-2M/(m_d+m_s)$. 
In this case, we have $c_1^\phi=c_1^{K^{*0}}=0$ and the $c_1/c_2$ term in Eq.~(\ref{eq:G rho00-1/3})  is vanishing, which is similar to the case of pseudoscalar-meson coalescence in Eq.~(\ref{c1-c2-meson}). 
However, since the coalescing particles, quarks on the one hand and mesons on the other hand, differ in their mass and quantum statistics, the results for the spin alignment are different in magnitude.
Nevertheless, the dependence on $p_T$, $\phi_p$, and $Y$ for quark coalescence with LS coupling is similar in shape as for the case of pseudoscalar-meson coalescence. 
Therefore, we do not present the numerical results for $\delta\rho^y_{00}(\mathbf{p})$ in this particular case. 

\begin{figure}
\begin{centering}
\includegraphics[scale=0.3]{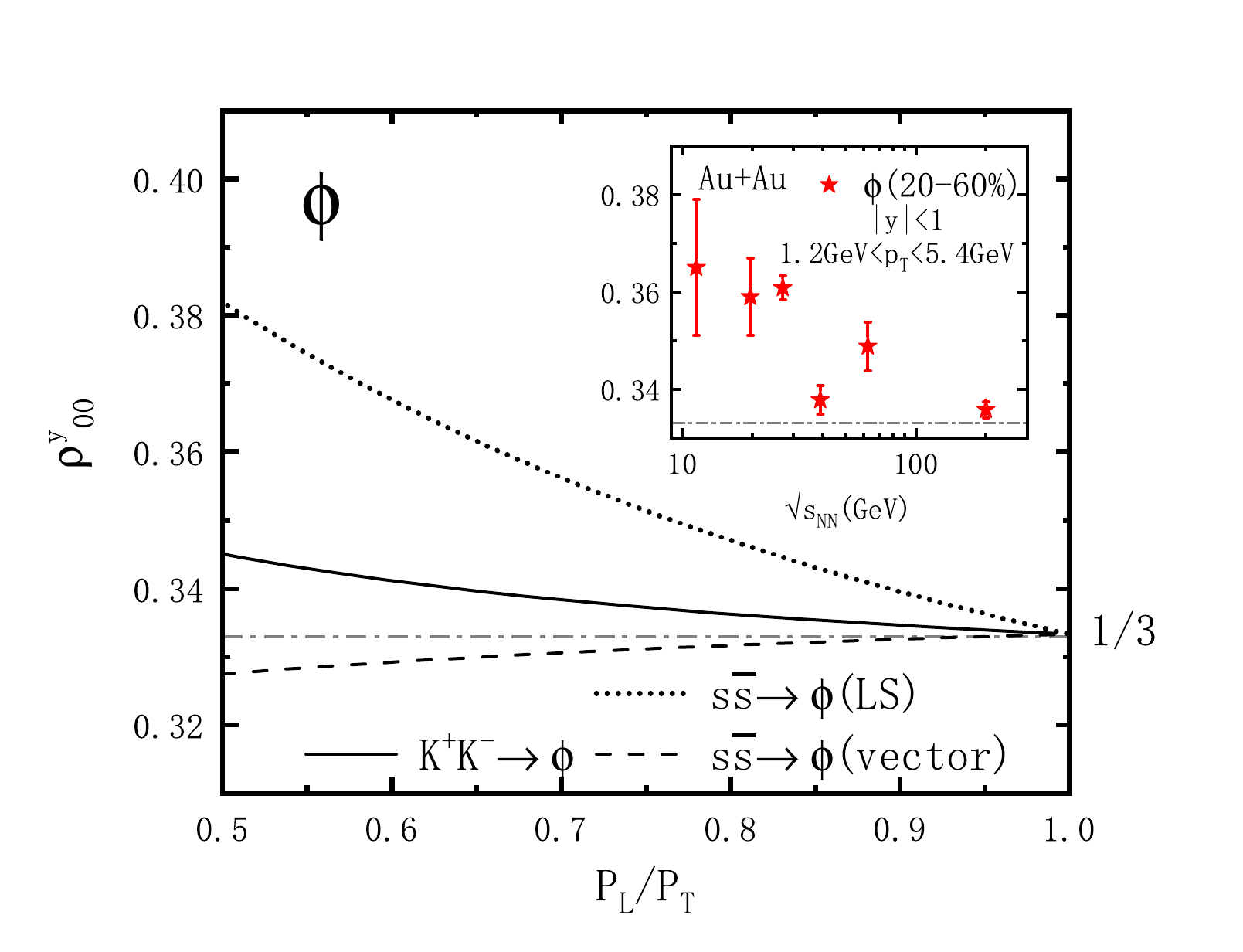}
\includegraphics[scale=0.3]{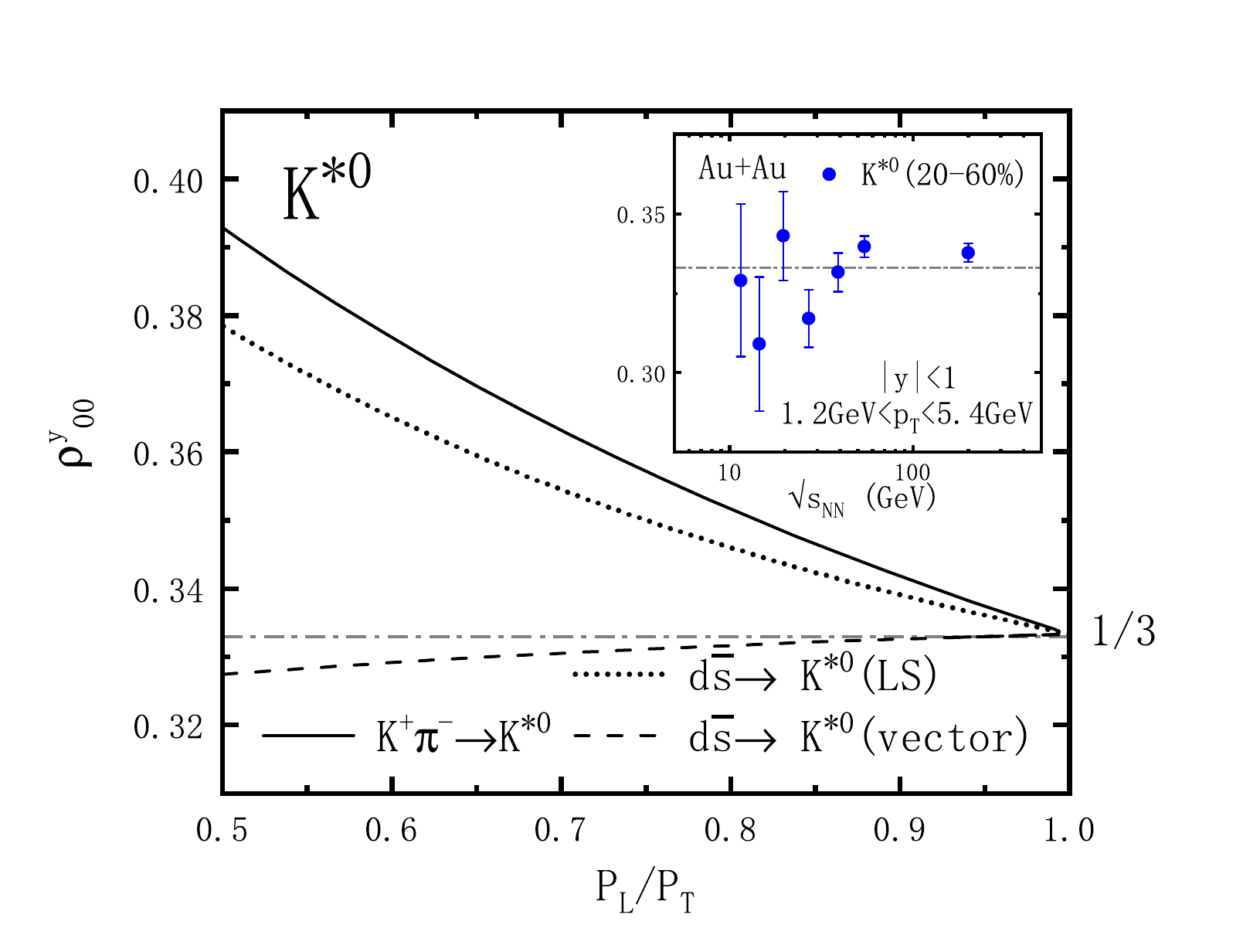}
\par\end{centering}
\caption{Numerical results for $\rho_{00}^y$ as functions of $P_{L}/P_{T}$ for vector mesons at ${\bf p}=0$. The solid lines represent the production via pseudoscalar-meson coalescence. The dotted and dashed lines represent the production via quark coalescence with bare vector coupling and  with LS coupling, respectively. We set $m_s=420$ MeV. The experimental data presented in the inset are taken from Ref.~\citep{STAR:2022fan} for the spin alignment in Au+Au collisions. 
\label{fig:rho00_PLPT}}
\end{figure}

In order to compare the results from pseudoscalar-meson and quark coalescence, let us assume that the rest frame of the vector meson coincides with the local rest frame of the colliding system. 
In this case, the spin alignment is given  $\delta \rho_{00}^{y}({\bf p}=0)$ and its dependence on $P_{L}/P_{T}$ can be calculated numerically. 
The results are shown in Fig.~\ref{fig:rho00_PLPT}. The dotted lines are calculated using the Lagrangian (\ref{eq:int-vector-tensor}) with $\zeta_\phi=-M/m_s$ and $\zeta_K=-2M/(m_d+m_s)$ such that only the LS-coupling term $\zeta_{\phi,K}\overset{\leftrightarrow}{\partial}^\mu$ contributes. 
In this case, $\delta \rho_{00}^y$ is positive.
If $\phi$ and $K^{*0}$ mesons are produced from quark coalescence with a bare vector coupling vertex, $\delta \rho_{00}^y$ will be negative, as indicated by the dashed lines. 
The results for the pseudoscalar-meson coalescence channel are shown by the solid lines and give positive  $\delta\rho_{00}^y$.
For $\phi$ mesons produced in this channel, the spin alignment is smaller than for the quark coalescence with pure LS coupling, while it is of the same order of magnitude or even slightly larger for $K^{*0}$ mesons. 
Due to interference between different production channels, the actually observed spin alignment will not be a linear sum of the three channels shown in Fig.~\ref{fig:rho00_PLPT}. 

Since $P_{L}/P_{T}$ increases with the collision energy, as estimated in Sec.~\ref{sec:anisotropic-hydrodynamics}, $\delta \rho_{00}^{y}$ for vector mesons produced via pseudoscalar-meson coalescence or quark coalescence decreases in magnitude with the collision energy, which is qualitatively consistent with the experimental data. 
Compared to $\phi$, $K^{*0}$ has a shorter life time and thus the detected $K^{*0}$ are more likely to be produced at a later time in collisions when the system is less anisotropic. 
This may explain why the experimentally observed spin alignment of $K^{*0}$ is consistent with zero, irrespective of the production channel.
On the other hand, only the quark coalescence with LS coupling can explain the magnitude of the experimentally observed spin alignment for $\phi$.
This is consistent with the picture that long-lived $\phi$ mesons are mainly produced in the early, and thus more anisotropic QGP stage of the collision. 

\section{Conclusions\label{sec: conclusion}}

In this paper, we study the spin alignment for $\phi$ and $K^{*0}$ mesons produced via coalescence at the quark as well as the hadron level. 
We demonstrate that the anisotropy of the particle distribution functions in heavy-ion collisions  induces a non-vanishing $\delta\rho_{00}^{y}$ for vector mesons, the sign and magnitude of which depend on the underlying production mechanism. 
For all production mechanism studied here, the magnitude of the spin alignment increases with the anisotropy.
If the vector meson is produced through quark coalescence without LS coupling, the anisotropy induces a negative $\delta\rho_{00}^{y}$ of order $\mathcal{O}(10^{-3})$. 
In contrast, the vector meson produced through quark coalescence with LS coupling and through pseudoscalar-meson coalescence has a positive $\delta\rho_{00}^{y}$ of order $\mathcal{O}(10^{-2})$. 
These values are comparable to those measured by experiments, which implies that the momentum anisotropy can in principle contribute to the observed spin alignment of vector mesons.
The measurement of the spin alignment thus provides a possible way to distinguish the production mechanism of vector mesons in heavy-ion collisions. 
More precise calculations require a more sophisticated dynamical model for the heavy-ion collision. 

Our framework can be further applied to studies of the spin alignment of heavy quarkonia (such as $J/\Psi$) at LHC energy and that of light vector mesons which are produced near their thresholds at RHIC-BES energies.
At LHC energy, the $c$ ($\bar{c}$) (anti-)quark is so heavy that its momentum can hardly change during its interaction with the medium. 
As a consequence, its momentum distribution remains highly anisotropic during the evolution of the system. 
This may generate a non-vanishing spin alignment of $J/\psi$. 
On the other hand, for light vector mesons produced near their thresholds, the hadronic reaction $NN\rightarrow NNV$ is the dominant channel. 
Since the momenta of the incoming nucleons are aligned along the beam direction, corresponding to an infinite $\xi$, a sizable spin alignment of these vector mesons is expected in collisions at $\sqrt{s_{NN}}<3\text{ GeV}$.

The current calculations are based on the RS distribution, which is a natural and straightforward method to include the momentum anisotropy. 
From the perspective of symmetry, a non-vanishing spin alignment must correspond to breaking the symmetry between the spin quantization direction and its perpendicular directions. 
In addition to the momentum anisotropy, other mechanisms, such as the vorticity field and the anisotropic shear-stress tensor field, can also break the symmetry and lead to the spin alignment of vector meson, which have been previously discussed in Refs.~\citep{Li:2022vmb,Dong:2023cng,Yang:2024fkn,Zhang:2024mhs}. 
These effects can be included in our framework by extending the production rate to next-to-leading order in the semi-classical expansion, at which order the gradients of the particle distributions contribute. 
The combined effect of the momentum anisotropy and vorticity (or shear-stress tensor) field will be discussed in future work.

\begin{acknowledgments}

W.B.D., Q.W. and Y.L.Y.~are supported by the National Natural Science Foundation of China under Grant No.~12135011. 
X.L.S.~is supported by the National Natural Science Foundation of China under Grant No.~12547102 and No.~12147101, and by the ExtreMe Matter Institute (EMMI) at the GSI Helmholtzzentrum fuer Schwerionenforschung GmbH, Darmstadt, Germany.
D.H.R.~is supported in part by the Deutsche Forschungsgemeinschaft (DFG, German Research Foundation) through the CRC-TR 211 `Strong-interaction matter under extreme conditions'' -- project number 315477589--TRR 211.
\end{acknowledgments}

\appendix

\bibliographystyle{h-physrev}
\bibliography{reference}

\end{document}